\def\Statusstring{Submitted to IEEE Transactions on Information Theory \\
                  January 14, 2006}

\documentclass[11pt,letter]{article}

\usepackage{amsmath,amscd,fancyhdr,graphics,pifont}
\usepackage{floatflt,subfigure,epsfig,moreverb,multicol,lscape}
\usepackage{supertabular,tabularx,multirow,bar,amssymb}
\usepackage{theorem,psfrag,afterpage,curves,epic, bbm}
\usepackage[english]{babel}

\newcommand{\ignore}[1]{}
\newcommand{\matr}[1]{\mathbf{#1}}
\newcommand{\vect}[1]{\mathbf{#1}}

\newcommand{\GF}[1]{\mathbb{F}_{#1}}

\newcommand{\defeq}{\triangleq}

\newcommand{\vu}{\vect{u}}
\newcommand{\vx}{\vect{x}}

\newcommand{\vy}{\vect{y}}

\newcommand{\HD}[2]{\mathcal{D}_{#1}^{(#2)}}
\newcommand{\iL}{\mathsf{L}}
\newcommand{\iX}{\mathsf{X}}

\newcommand{\us}{\upsilon} % unerased symbols
\newcommand{\usSet}{\Upsilon} % unerased symbols

\makeatletter
  \def\revddots{\mathinner{\mkern1mu\raise\p@
  \vbox{\kern7\p@\hbox{.}}\mkern2mu
  \raise4\p@\hbox{.}\mkern2mu\raise7\p@\hbox{.}\mkern1mu}}
  \makeatother

\newtheorem{Lemma}{Lemma}
\newtheorem{Theorem}[Lemma]{Theorem}

\newtheorem{Corollary}[Lemma]{Corollary}

\theoremstyle{plain}

\newtheorem{PreDefinition}[Lemma]{{\textbf{Definition}}}
  \newenvironment{Definition}%
    {\begin{PreDefinition}}{\hfill$\square$\end{PreDefinition}}

\theoremstyle{plain}

\newtheorem{PreAlgorithm}[Lemma]{{\textbf{Algorithm}}}
    {\begin{PreAlgorithm}\upshape}{\hfill$\square$\end{PreAlgorithm}}

\newtheorem{PreRemark}[Lemma]{{\textbf{Remark}}}
    {\begin{PreRemark}\upshape}{\hfill$\square$\end{PreRemark}}

\newtheorem{PreExample}[Lemma]{{\textbf{Example}}}
  \newenvironment{Example}%
    {\begin{PreExample}\upshape}{\hfill$\square$\end{PreExample}}

\newenvironment{Proof}%
  {\noindent \emph{Proof:}}{\hfill$\square$}

\begin{document}

\title{On the Existence of Universally Decodable Matrices%
  \footnote{The first author was supported by NSF Grant TF-0514801.
  The second author was supported by NSF Grants ATM-0296033 and DOE
            SciDAC and by ONR Grant N00014-00-1-0966.  This work was presented in part
            at the 43rd Annual Allerton Conference, Monticello, IL,
            September 2005~\cite{Ganesan:Boston:05:1}.}}

\author{Ashwin Ganesan%
  \thanks{ECE Department, University of
          Wisconsin-Madison, 1415 Engineering Drive Madison, WI 53706, USA.
          Email: \texttt{ganesan@cae.wisc.edu}. A.G.~is the corresponding author.}
  \ and Pascal O.~Vontobel%
  \thanks{Was with ECE Department, University of
          Wisconsin-Madison, 1415 Engineering Drive Madison, WI 53706, USA.
          Email: \texttt{pascal.vontobel@ieee.org}.}}

\date{}

\maketitle

\vspace{-6cm}
\begin{flushright}
  \texttt{\Statusstring}\\[1cm]
\end{flushright}
\vspace{+4cm}

\begin{abstract}
  Universally decodable matrices (UDMs) can be used for coding purposes when
  transmitting over slow fading channels.  These matrices are parameterized by
  positive integers $L$ and $N$ and a prime power $q$.  The main result of
  this paper is that the simple condition $L \leq q+1$ is both necessary and
  sufficient for $(L,N,q)$-UDMs to exist.  The existence proof is constructive
  and yields a coding scheme that is equivalent to a class of
  codes that was proposed
  by Rosenbloom and Tsfasman. Our work resolves an open problem posed recently
  in the literature.

\end{abstract}


\newpage

\section{Introduction}
\label{sec:introduction:1}

Let $L$ and $N$ be positive integers, let $q$ be a prime power, let
$[M]
\defeq \{ 0, \ldots, M-1 \}$ for any positive integer $M$, and let $[M] \defeq
\{ \ \}$ for any non-positive integer $M$. While studying slow
fading channels, Tavildar and
Viswanath~\cite{Tavildar:Viswanath:05:1},\cite{Tavildar:Viswanath:05:2}
introduced a communication system which works as follows. An
information (column) vector $\vect{u} \in \GF{q}^N$ is encoded into
codeword vectors $\vx_{\ell}
\defeq \matr{A}_{\ell} \cdot \vect{u} \in \GF{q}^{N}$, $\ell \in
[L]$, where $\matr{A}_0, \ldots, \matr{A}_{L-1}$ are $L$ matrices
over $\GF{q}$ and of size $N \times N$.  Upon sending $\vx_{\ell}$
over the $\ell$-th channel we receive $\vy_{\ell} \in (\GF{q} \cup
\{ ? \})^N$, where the question mark denotes an erasure. The
channels are such that the received vectors $\vect{y}_0, \ldots,
\vect{y}_{L-1}$ can be characterized as follows: there are integers
$\us_0, \ldots, \us_{L-1}$ (that can vary from transmission to transmission),
$0 \leq \us_{\ell} \leq N$, $\ell \in [L]$, such that the first $\us_{\ell}$
entries of $\vect{y}_{\ell}$ are non-erased and agree with the corresponding
entries of $\vect{x}_{\ell}$ and such that the last $N - \us_{\ell}$ entries
of $\vect{y}_{\ell}$ are erased.

Based on the non-erased entries we would like to reconstruct
$\vect{u}$. The obvious decoding approach works as follows:
construct an $N \times N$-matrix $\matr{A}$ that stacks the $\us_0$
first rows of $\matr{A}_0$, $\ldots$, the $\us_{L-1}$ first rows of
$\matr{A}_{L-1}$.   Since $\vect{u}$ is arbitrary in $\GF{q}^N$, a
necessary condition for successful decoding is that $\sum_{\ell \in
[L]} \us_{\ell} \geq N$. Because we would like to be able to decode
successfully for all $L$-tuples $(\us_0, \ldots, \us_{L-1})$ that
satisfy this necessary condition, we must guarantee that the matrix
$\matr{A}$ has full rank for all possible $L$-tuples $(\us_0,
\ldots, \us_{L-1})$ with $\sum_{\ell \in [L]} \us_{\ell} = N$.
 This will automatically also guarantee that the matrix $\matr{A}$
 has full rank for all possible $L$-tuples $(\us_0,
\ldots, \us_{L-1})$ with $\sum_{\ell \in [L]} \us_{\ell} \geq N$.
Matrices that fulfill this condition are called
$(L,N,q)$-~universally decodable matrices (UDMs).  A more precise
definition and some examples are given in the next section.  It is
trivially verified that $(L,N,q)$-UDMs always exist when $N=1$, so
throughout the rest of this paper it is assumed that $N \ge 2$.

Given the definition of $(L,N,q)$~-UDMs, there are several immediate
questions. For what values of $L$, $N$, and $q$ do such matrices
exist? What are the properties of these matrices? How can one
construct such matrices?  The authors in
\cite{Tavildar:Viswanath:05:1} provide a construction for these
matrices for the cases $L = 3$, any $N$, and $q = 2$. For the case
$L=4$, any $N$, and $q=3$,
Doshi~\cite{Doshi:05:1},\cite{Tavildar:Viswanath:05:1} conjectures
that a particular construction yields $(L{=}4,N,q{=}3)$-UDMs. The
existence and construction of $(L,N,q)$-UDMs for the general case is
proposed as an open problem in \cite{Tavildar:Viswanath:05:1}.  This
is the take-off point for the present paper. Ganesan and
Boston~\cite{Ganesan:Boston:05:1} showed that a necessary condition
for $(L,N,q)$-UDMs to exist is $L \le q+1$ and conjectured that this
condition is also sufficient.  In this paper we complete the
resolution of this problem on the existence of $(L,N,q)$-UDMs.

The paper is structured as follows. In Sec.~\ref{sec:udm:1} we
properly define UDMs and prove the necessary condition.
Sec.~\ref{sec:explicit:construction:udms:1} discusses the explicit
construction that proves the sufficiency part and
Sec.~\ref{sec:conclusions:1} contains some concluding remarks.
Sec.~\ref{sec:proof:cor:theorem:udms:construction:1} contains the
proof of Cor.~\ref{cor:theorem:udms:construction:1}, and
Sec.~\ref{sec:hasse:derivatives:1} collects some results on Hasse
derivatives which are the main tool for the proof of our UDMs
construction.

\section{Universally Decodable Matrices}
\label{sec:udm:1}

The notion of universally decodable matrices (UDMs) was introduced
by Tavildar and Viswanath~\cite{Tavildar:Viswanath:05:1}. Before we
give the definition of UDMs, let us agree on some notation. For any
positive integer $N$, we let $\matr{I}_N$ be the $N \times N$
identity matrix, and we let $\matr{J}_N$ be the $N \times N$ matrix
where all entries are zero except for the anti-diagonal entries that
are equal to one; i.e., $\matr{J}_N$ contains the rows of
$\matr{I}_N$ in reverse order.  For any positive integer $L$ and any
non-negative integer $N$ we define the set
\begin{align*}
  \usSet^{= N}_{L}
    &\defeq
       \left\{
           (\us_0, \ldots, \us_{L-1})
         \
         \left|
         \
           0 \leq \us_{\ell} \leq N,
           \ell \in [L], \
           \sum_{\ell \in [L]} \us_{\ell} = N
         \right.
       \right\}.
\end{align*}
Throughout the rest of this paper, indices start at $0$ and not at
$1$. We let $[\matr{M}]_{n,k}$ denote the $(n,k)$-th entry of the
matrix $\matr{M}$ and we let $[\vect{v}]_{n}$ denote the $n$-th
entry of the vector $\vect{v}$.

\begin{Definition}
  \label{def:udms:1}

  Let $N$ and $L$ be some positive integers and let $q$ be a prime power. The
  $L$ matrices $\matr{A}_0, \ldots, \matr{A}_{L-1}$ over $\GF{q}$ and of size
  $N \times N$ are $(L,N,q)$-UDMs, or simply UDMs, if for every partition of
  $N$ into $L$ non-negative summands $(\us_0,\us_1,\ldots,\us_{L-1}) \in
  \usSet^{= N}_{L}$, the following condition is satisfied: the $N \times N$
  matrix composed of the first $\us_0$ rows of $\matr{A}_0$, the first $\us_1$
  rows of $\matr{A}_1$, $\ldots$, the first $\us_{L-1}$ rows of
  $\matr{A}_{L-1}$ has full rank.  We call this condition the {\em UDMs
  condition}.
\end{Definition}

\begin{Example}
  \label{ex:udm:1}

  Let $N$ be any positive integer, let $q$ be any prime power, and let $L
  \defeq 2$. Let $\matr{A}_0 \defeq \matr{I}_N$ and let $\matr{A}_1 \defeq
  \matr{J}_N$. It can easily be checked that $\matr{A}_0, \matr{A}_1$ are
  $(L{=}2,N,q)$-UDMs. Indeed, let for example $N \defeq 5$. We must check
  that for any non-negative integers $\us_1$ and $\us_2$ such that $\us_1 +
  \us_2 = 5$ the UDMs condition is fulfilled. E.g.~for $(\us_1, \us_2) = (3,
  2)$ we must show that the matrix
  \begin{align*}
    \begin{pmatrix}
      1 & 0 & 0 & 0 & 0 \\
      0 & 1 & 0 & 0 & 0 \\
      0 & 0 & 1 & 0 & 0 \\
      0 & 0 & 0 & 0 & 1 \\
      0 & 0 & 0 & 1 & 0
    \end{pmatrix}
  \end{align*}
  has rank $5$, which can easily be verified.
\end{Example}

\begin{Example}
  \label{ex:udm:2}

  In order to give the reader a feeling how UDMs might look like for $L > 2$,
  we give here a simple example for $L = 4$, $N = 3$ and $q = 3$, namely
  \begin{align*}
    \matr{A}_0
      &= \begin{pmatrix}
           1 & 0 & 0 \\
           0 & 1 & 0 \\
           0 & 0 & 1
         \end{pmatrix},
         \quad
    \matr{A}_1
       =
         \begin{pmatrix}
           0 & 0 & 1 \\
           0 & 1 & 0 \\
           1 & 0 & 0
         \end{pmatrix},
         \quad
    \matr{A}_2
       =
         \begin{pmatrix}
           1 & 1 & 1 \\
           0 & 1 & 2 \\
           0 & 0 & 1
         \end{pmatrix},
         \quad
    \matr{A}_3
       =
         \begin{pmatrix}
           1 & 2 & 1 \\
           0 & 1 & 1 \\
           0 & 0 & 1
         \end{pmatrix}.
  \end{align*}
  One can verify that for all $(\us_0, \us_1, \us_2, \us_3)$ such that $\us_0+
  \us_1+ \us_2+ \us_3=4$ (there are $20$ such four-tuples) the UDMs condition
  is fulfilled and hence the above matrices are indeed UDMs. For example, for
  $(\us_0,\us_1,\us_2,\us_3) = (0,0,3,0)$, $(\us_0,\us_1,\us_2,\us_3) =
  (0,0,1,2)$, and $(\us_0,\us_1,\us_2,\us_3) = (1,1,0,1)$ the UDMs condition
  means that we have to check if the matrices
  \begin{align*}
    \begin{pmatrix}
      1 & 1 & 1 \\
      0 & 1 & 2 \\
      0 & 0 & 1
    \end{pmatrix},
    \quad
    \begin{pmatrix}
      1 & 1 & 1 \\
      1 & 2 & 1 \\
      0 & 1 & 1
    \end{pmatrix},
    \quad
    \begin{pmatrix}
      1 & 0 & 0 \\
      0 & 0 & 1 \\
      1 & 2 & 1
    \end{pmatrix}
  \end{align*}
  have rank $3$, respectively, which is indeed the case. Before concluding
  this example, let us remark that the above UDMs are the same UDMs that
  appeared in~\cite{Doshi:05:1}
  and~\cite[Sec.~4.5.4]{Tavildar:Viswanath:05:1}.
\end{Example}

\begin{Theorem}
  \label{lemma:maximal:L:1}

  If $N \geq 2$ then a necessary condition for $(L,N,q)$-UDMs to exist is $L
  \leq q + 1$.

\end{Theorem}

\begin{Proof}
  We use the notation $\matr{A}_i[j]$ to denote the $j$-th row of
  $\matr{A}_i$, with $j \in [N]$.  Suppose the $N \times N$ matrices
  $\matr{A}_0,\ldots,\matr{A}_{L-1}$ are UDMs. Let $S_{N-2}$ be the $(N-2)$
  dimensional subspace spanned by
  $\{\matr{A}_0[0],\matr{A}_0[1],\ldots,\matr{A}_0[N-3]\}$. The quotient space
  $Q\defeq\GF{q}^{N} / S_{N-2}$ is isomorphic to a 2-dimensional subspace of
  $\GF{q}^{N}$. Given the set of $L$ vectors $\mathcal{A} \defeq
  \{\matr{A}_0[N-2], \matr{A}_1[0], \matr{A}_2[0], \ldots,
  \matr{A}_{L-1}[0]\}$, form the corresponding set of cosets $C \defeq \{a +
  S_{N-2} ~|~ a \in \mathcal{A}\}$ in $Q$. Now by the UDMs condition with
  $\us_0=N-2$, any two elements of $C$ must be linearly independent. Thus, if
  a particular coset is in $C$, $q-2$ other cosets from $Q$ cannot be in $C$.
  So we partition $Q$ into sets containing $q-1$ cosets, with the $q-1$ cosets
  in each partition being linearly dependent and with at most one coset from a
  partition being a member of $C$. The cardinality of $Q \setminus \{ \vect{0}
  \}$ is $q^2-1$. We obtain the necessary condition on the number of
  partitions
  \begin{align*}
    \#(C)
      &\le \frac{q^2-1}{q-1}.
  \end{align*}
  But $L = \#(C)$, and so the theorem follows.
\end{Proof}

\mbox{}

As mentioned earlier, it is assumed throughout this paper that $N \geq 2$.  In
\cite{Tavildar:Viswanath:05:1}, a construction was given for
$(L{=}3,N,q{=}2)$-UDMs, and a construction was conjectured to yield
$(L{=}4,N,q{=}3)$-UDMs.  As a corollary of the theorem above, we obtain that
there do not exist any UDMs for $q=2, L \ge 4$, and that there do not exist
any UDMs for $q=3, L \ge 5$.

\section{An Explicit Construction of UDMs}
\label{sec:explicit:construction:udms:1}

In this section we present an explicit construction of $(L,N,q)$-UDMs when $L
\leq q+1$, cf.~Th.~\ref{theorem:udms:construction:1} and
\cite{Vontobel:Ganesan:05:1}.  Before we proceed, we need some
definitions. First, whenever necessary we use the natural mapping of
the integers into the prime subfield\footnote{When $q = p^s$ for
some prime $p$ and some positive integer $s$ then $\GF{p}$ is a
subfield of $\GF{q}$ and is called the prime subfield of $\GF{q}$.
$\GF{p}$ can be identified with the integers where addition and
multiplication are modulo $p$.} of $\GF{q}$. Secondly, we define the
binomial coefficient ${a \choose b}$ in the usual way. Note that ${a
\choose b} = 0$ for all $a < b$.

\begin{Definition}
  \label{def:taylor:polynomial:expansion:1}

  Let $a(\iX) \defeq \sum_{k=0}^{d} a_k \iX^k \in \GF{q}[\iX]$ be a polynomial
  and let $\beta \in \GF{q}$. The Taylor polynomial expansion of $a(\iX)$
  around $\iX = \beta$ is defined to be $a(\iX) = \sum_{n=0}^{d} a_{\beta,n}
  (\iX - \beta)^n \in \GF{q}[\iX]$ for suitably chosen $a_{\beta,n} \in
  \GF{q}$, $0 \leq n \leq d$, such that equality holds.
\end{Definition}

It can be verified that the Taylor polynomial coefficients $a_{\beta,n}$ can
be expressed using Hasse derivatives\footnote{See
Sec.~\ref{sec:hasse:derivatives:1} for the definition and some properties of
Hasse derivatives.} of $a(\iX)$, i.e.~$a_{\beta,n} = a^{(n)}(\beta) =
\sum_{k=0}^{d} a_k {k \choose n} \beta^{k-n}$. On the other hand, the
coefficients of $a(\iX)$ can be expressed as $a_k = \sum_{n=0}^{d} a_{\beta,n}
{n \choose k} (-\beta)^{n-k}$.

\begin{Lemma}
  \label{lemma:zero:multiplicity:1}

  Let $a(\iX) \defeq \sum_{k=0}^{d} a_k \iX^k \in \GF{q}[\iX]$ be a non-zero
  polynomial, let $\beta \in \GF{q}$, and let $a(\iX) = \sum_{n=0}^{d}
  a_{\beta,n} (\iX - \beta)^n \in \GF{q}[\iX]$ be the Taylor polynomial
  expansion of $a(\iX)$ around $\iX = \beta$. The polynomial $a(\iX)$ has a
  zero at $\iX = \beta$ of multiplicity $m$ if and only if $a_{\beta,n} = 0$
  for $0 \leq n < m$ and $a_{\beta,m} \neq 0$.
\end{Lemma}

\begin{Proof}
  Obvious.
\end{Proof}

\mbox{}

In the following, evaluating the $n$-th Hasse derivative
$u^{(n)}(\iL)$ of a polynomial $u(\iL)$ at $\iL = \infty$ shall
result in the value $u_{N-1-n}$, i.e.~we set $u^{(n)}(\infty) \defeq
u_{N-1-n}$.

\begin{Theorem}
  \label{theorem:udms:construction:1}

  Let $N$ and $L$ be positive integers, let $q$ be some prime power.  If $L
  \leq q+1$ then the following $L$ matrices over $\GF{q}$ of size $N \times N$
  are $(L,N,q)$-UDMs:
  \begin{align*}
    &
    \matr{A}_0
       \defeq \matr{I}_{N},
    \quad
    \matr{A}_1
       \defeq \matr{J}_{N},
    \quad
    \matr{A}_{2},
    \quad
    \ldots,
    \quad
    \matr{A}_{L-1}, \\
    &
    \text{ where } \
    [\matr{A}_{\ell+2}]_{n,k}
       \defeq {k \choose n} \alpha^{\ell (k-n)}, \
              (\ell,n,k) \in [L-2] \times [N] \times [N],
  \end{align*}
  where $\alpha$ is any primitive element in $\GF{q}$.  Note that ${k \choose
  n}$ is to be understood as follows: compute ${k \choose n}$ over the
  integers and apply only then the natural mapping to $\GF{q}$.
\end{Theorem}

\begin{Proof}
  Follows easily from
  Cor.~\ref{cor:theorem:udms:construction:1} and its proof, together with the
  paragraph after Def.~\ref{def:taylor:polynomial:expansion:1}.
\end{Proof}

\begin{Corollary}
  \label{cor:theorem:udms:construction:1}

  Let us associate the information polynomial $u(\iL) \defeq \sum_{k \in [N]}
  u_k \iL^k \in \GF{q}[\iL]$ with $u_k \defeq [\vu]_k$, $k \in [N]$, to the
  information vector $\vu$. The construction in the above theorem results in a
  coding scheme where the vector $\vu$ is mapped to the vectors $\vx_0,
  \ldots, \vx_{L-1}$ with entries
  \begin{align*}
    [\vx_{\ell}]_n
      &= u^{(n)}(\beta_{\ell}),
         \quad (\ell,n) \in [L] \times [N],
  \end{align*}
  where $\beta_0 \defeq 0$, $\beta_1 \defeq \infty$, $\beta_{\ell+2} \defeq
  \alpha^{\ell}$, $\ell \in [L-2]$. (Note that because $\alpha$ is a primitive
  element of $\GF{q}$, all $\beta_{\ell}$, $\ell \in [L]$, are distinct.) This
  means that over the $\ell$-th channel we are transmitting the coefficients
  of the Taylor polynomial expansion of $u(\iL)$ around $\iL = \beta_{\ell}$.
\end{Corollary}

\begin{Proof}
  See Sec.~\ref{sec:proof:cor:theorem:udms:construction:1}.
\end{Proof}

\mbox{}

As already mentioned in Sec.~\ref{sec:introduction:1}, the construction of
UDMs in Th.~\ref{theorem:udms:construction:1}~/~%
Cor.~\ref{cor:theorem:udms:construction:1} is essentially equivalent to a
class of codes presented by Rosenbloom and
Tsfasman~\cite{Rosenbloom:Tsfasman:97:1}.\footnote{Note that the communication
system mentioned in Sec.~1 of~\cite{Rosenbloom:Tsfasman:97:1} also talks about
parallel channels: however, that communication system would correspond to (in
our notation) sending $L$ symbols over $N$ channels. On the other hand, the
communication system that is mentioned in Nielsen~\cite[Ex.~18]{Nielsen:00:1}
is more along the lines of the Tavildar-Viswanath channel
model~\cite{Tavildar:Viswanath:05:1} mentioned in
Sec.~\ref{sec:introduction:1}.}

\section{Concluding Remarks}
\label{sec:conclusions:1}

We have shown that $(L,N,q)$-UDMs exist if and only if $L \leq q+1$,
and the existence proof is constructive.  This completely resolves
the open problem posed in~\cite{Tavildar:Viswanath:05:1} on the
existence of $(L,N,q)$-UDMs.  Many open questions remain.  It is of
interest to determine whether there are also other UDM constructions
that are not simply reformulations of the present UDMs, and to
obtain efficient decoding algorithms that exploit the structure of
these matrices.  Recent developments in these directions - on the
uniqueness of the present constructions, how to efficiently decode,
and the resolution of a conjecture in \cite{Tavildar:Viswanath:05:1}
- will be presented at \cite{Vontobel:Ganesan:06:1} and in a
forthcoming paper.

\appendix

\section{Proof of Corollary~\ref{cor:theorem:udms:construction:1}}

\label{sec:proof:cor:theorem:udms:construction:1}

We have to check the UDMs condition for all $(\us_0, \ldots,
\us_{L-1}) \in \usSet^{= N}_{L}$. Fix such a tuple $(\us_0, \ldots,
\us_{L-1}) \in \usSet^{=N}_{L}$ and let $\psi$ be the mapping of the
vector $\vect{u}$ to the non-erased entries of the vectors
$\vect{y}_{\ell}$, $\ell \in [L]$; it is clear that $\psi$ is a
linear mapping. Reconstructing $\vect{u}$ is therefore nothing else
than applying the mapping $\psi^{-1}$ to the non-erased positions of
$\vect{y}_{\ell}$, $\ell \in [L]$. However, this gives a unique
vector $\vect{u}$ only if $\psi$ is an injective function. Because
$\psi$ is linear, showing injectivity of $\psi$ is equivalent to
showing that the kernel of $\psi$ contains only the vector $\vect{u}
= \vect{0}$, or equivalently, only the polynomial $u(\iL) = 0$.

So, let us show that the only possible pre-image of
\begin{align*}
  [\vect{y}_{\ell}]_n
    &= 0
       \quad
       (\ell \in [L], \ n \in [\us_{\ell}])
\end{align*}
or, equivalently, of
\begin{align*}
  [\vect{x}_{\ell}]_n
    &= 0
       \quad
       (\ell \in [L], \ n \in [\us_{\ell}])
\end{align*}
is $u(\iL) = 0$. Using the definition of $[\vect{x}_{\ell}]_n$ this is
equivalent to showing that
\begin{alignat}{2}
  u^{(n)}(\beta_{\ell})
    &= 0
       \quad
       &&(\ell \in [L] \setminus \{ 1 \}, \ n \in [\us_{\ell}])
         \label{eq:zero:hasse:derivatives:1} \\
  u_{N-1-n}
     = u^{(n)}(\beta_{\ell})
    &= 0
       \quad
       && (\ell = 1, \ n \in [\us_{\ell}])
         \label{eq:u:zero:positions:1}
\end{alignat}
implies that $u(\iL) = 0$. In a first step,
Eq.~\eqref{eq:zero:hasse:derivatives:1} and
Lemma~\ref{lemma:zero:multiplicity:1} tell us that $\beta_{\ell}$,
$\ell \in [L] \setminus \{ 1 \}$, must be a root of $u(\iL)$ of
multiplicity at least $\us_{\ell}$. Using the fundamental theorem of
algebra we get
\begin{align}
  \deg(u(\iL))
    &\geq
       \sum_{\ell \in [L] \setminus \{ 1 \}}
         \us_\ell
     = N - \us_1
  \quad\quad
  \text{ or }
  \quad\quad
  u(\iL) = 0.
         \label{eq:lower:bound:number:of:roots:1}
\end{align}
In a second step, Eq.~\eqref{eq:u:zero:positions:1} tells us that we
must have $\deg(u(\iL)) \leq N - 1 - \us_1$. Combining this
with~\eqref{eq:lower:bound:number:of:roots:1}, we obtain the desired
result that $u(\iL) = 0$.

\section{Hasse Derivatives}

\label{sec:hasse:derivatives:1}

Hasse derivatives were introduced in~\cite{Hasse:36:1}. Throughout this
appendix, let $q$ be some prime power. For any non-negative integer $i$, the
$i$-th Hasse derivative of a polynomial $a(\iX) \defeq \sum_{k=0}^{d} a_k
\iX^k \in \GF{q}[\iX]$ is defined to be\footnote{The $i$-th formal
derivative equals $i!$ times the Hasse derivative: so, for fields with
characteristic zero there is not a big difference between these two
derivatives since $i!$ is always non-zero, however for finite fields there can
be quite a gap between these two derivatives since $i!$ can be zero or
non-zero.}
\begin{align*}
  a^{(i)}(\iX)
    &\defeq
       \HD{\iX}{i}
         \left(
           \sum_{k=0}^{d}
             a_k \iX^k
         \right)
     \defeq
       \sum_{k=0}^{d}
         {k \choose i}  a_k \iX^{k-i}.
\end{align*}
Note that when $i > k$ then ${k \choose i} \iX^{k-i} = 0$, i.e.~the zero
polynomial. Be careful that $\HD{\iX}{i_1} \HD{\iX}{i_2} \neq
\HD{\iX}{i_1 + i_2}$ in general. However, it holds that $\HD{\iX}{i_1}
\HD{\iX}{i_2} = {i_1 + i_2 \choose i_1} \HD{\iX}{i_1 + i_2}$.

We list some well-know properties of the Hasse derivatives:
\begin{align}
  \HD{\iX}{i}
    \big(
      \gamma f(\iX) + \eta g(\iX)
    \big)
      &= \gamma \HD{\iX}{i}\big( f(\iX) \big)
         +
         \eta \HD{\iX}{i}\big( g(\iX) \big),
           \nonumber \\
  \HD{\iX}{i}
    \big(
      f(\iX) g(\iX)
    \big)
      &= \sum_{i'=0}^{i}
           \HD{\iX}{i'}\big( f(\iX) \big)
           \HD{\iX}{i-i'}\big( g(\iX) \big),
             \nonumber \\
  \HD{\iX}{i}
    \left(
      \prod_{h \in [M]}
        f_{h}(\iX)
    \right)
      &= \sum_{(i_0, \ldots, i_{M-1}) \in \usSet^{= i}_{M}} \
           \prod_{h \in [M]}
             \HD{\iX}{i_{h}}\big( f_{h}(\iX) \big),
               \label{eq:hasse:derivative:property:1:1} \\
  \HD{\iX}{i}
    \left(
      (X-\gamma)^k
    \right)
      &= {k \choose i} (X-\gamma)^{k-i},
           \label{eq:hasse:derivative:property:1:2}
\end{align}
where $k$ and $i$ are some non-negative integers, $M$ is some
positive integer, and where $\gamma, \eta \in \GF{q}$. The fact that
a $\usSet$-set appears in Def.~\ref{def:udms:1} and in
Eq.~\eqref{eq:hasse:derivative:property:1:1} certainly points
towards the usefulness of Hasse derivatives for constructing UDMs.

\section*{Acknowledgments}

We would like to thank Richard Brualdi for early discussions on this
problem, Ralf Koetter for pointing out to us the
papers~\cite{Rosenbloom:Tsfasman:97:1, Nielsen:00:1}, Kamil
Zigangirov for providing us with a copy
of~\cite{Rosenbloom:Tsfasman:97:1}, and Nigel Boston for general
discussions on this work.

{
\bibliographystyle{ieeetr}
\bibliography{references}
}
\end{document}